\documentstyle[NATO,psfig]{crckapb} 

\begin{opening}
\title{K CORRECTIONS FOR TYPE IA SUPERNOVAE AND \protect \\
A TEST FOR SPATIAL VARIATION OF THE HUBBLE CONSTANT}
\subtitle{The Supernova Cosmology Project: II}

\author{A. Kim$^1$}
\author{S. Deustua}
\author{S. Gabi}
\author{G. Goldhaber}
\author{D. Groom}
\author{I. Hook}
\author{M. Kim}
\author{J. Lee}
\author{R. Pain$^2$}
\author{C. Pennypacker}
\author{S. Perlmutter}
\author{I. Small}
\institute{E. O. Lawrence Berkeley National Laboratory \& Center
for Particle Astrophysics,
University of California, Berkeley}
\author{A. Goobar}
\institute{University of Stockholm}
\author{R. Ellis}
\author{K. Glazebrook$^3$}
\author{R. McMahon}
\institute{Institute of Astronomy, Cambridge University}
\author{B. Boyle}
\author{P. Bunclark}
\author{D. Carter}
\author{M. Irwin}
\institute{Royal Greenwich Observatory}
\author{H. Newberg}
\institute{Fermi National Accelerator Laboratory}
\author{A. V. Filippenko}
\author{T. Matheson}
\institute{University of California, Berkeley}
\author{M. Dopita}
\author{J. Mould}
\institute{MSSSO, Australian National University}
\author{W. Couch}
\institute{University of New South Wales}

\end{opening}

\runningtitle{K CORRECTIONS}

\begin{document}
\footnotetext[1]{Presented by A. Kim, e-mail address:{\it agkim@LBL.gov}}
\footnotetext[2]{Current address:  CNRS-IN2P3, Paris}
\footnotetext[3]{Current address: Anglo-Australian Observatory}

\begin{abstract}

Cross-filter K corrections for a sample of ``normal'' Type Ia supernovae
(SNe)
have been calculated for a range of epochs.
With appropriate filter choices, the combined
statistical and systematic K correction dispersion of the full
sample lies within 0.05 mag for redshifts $z<0.7$.  This narrow dispersion
of the calculated K correction allows the Type Ia
to be used as a cosmological probe.  We use the K corrections with
observations
of seven SNe at redshifts $0.3 < z <0.5$ to bound the possible difference
between the locally measured Hubble constant ($H_L$) and the true cosmological
Hubble constant ($H_0$).

\end{abstract}
\section{Introduction}

The homogeneity and brightness
of Type Ia SN peak magnitudes have long made it
a popular standard candle (Branch and Tammann 1992\nocite{br:araa}). 
Recent observations, as discussed in this meeting, now indicate that
Type Ia's appear to form a family, rather
than a set of identical objects.  However, magnitude corrections based
on independent observables can
make the Type Ia a
calibrated candle.
Correlations between peak magnitude
and light curve shape
(Hamuy et al. 1995\nocite{ha:hubble};
Riess, Press, \& Kirshner 1995\nocite{re:lcs}) or spectral features
(Nugent et al. 1995\nocite{nu:1995}), have
made it possible to use the Type Ia as
a calibrated candle with $B$ magnitude dispersions of $\sim 0.2$ mag.
Alternatively, the use of Type Ia sub-samples
can provide an improved standard candle.
Supernovae with high quality data that pass certain color cuts
show a low dispersion in $B$ magnitudes of $< 0.28$
mag (Vaughan et al. 1995\nocite{va:inp}).

In order to use SNe as a tool for cosmology, we use K corrections
(Oke \& Sandage 1968)\nocite{ok:kcorr} to account for the
redshifting of spectra and its effect on nearby and distant flux
measurements
in different passbands. 
The sparseness of
spectroscopic observations of any individual
high-redshift SN requires us to use a statistical
approach to K corrections.
K corrections must therefore show a
small magnitude dispersion to maintain the usefulness of the SN 
as a standard or calibrated candle.

As of June 1995, the Supernova Cosmology Project
had discovered seven SNe lying in the range $0.3<z<0.5$
(Perlmutter et al. 1995; Perlmutter et al. 1996)
\nocite{sn1992bi}.  In a
standard Friedmann cosmology, their apparent magnitudes
depend on $\Omega$ and $\Lambda$ through the luminosity distance.
Perturbations on a homogeneous and isotropic
universe can cause the locally measured Hubble constant ($H_L$) to differ
from the global Hubble constant ($H_0$), as demonstrated by Turner, Cen, and
Ostriker (1992)\nocite{tu:hubble}.
In this case
if the SN calibrators lie within the local peculiar flow, then the
high-z SN
apparent magnitudes will additionally depend on $H_L/H_0$.
A scenario in which $H_L/H_0 > 1$ has been suggested by Bartlett et al. (1994)
\nocite{ba:h30} and others
to reconcile theoretical arguments for a low Hubble
constant with the recent observational
evidence for a large Hubble constant.
We use our seven supernovae, with K corrections, to place bounds
on $H_L/H_0$.

\section{The K Correction}

The generalization of the K correction of Oke \& Sandage relating
$y$ band observations of redshifted objects with $x$ band observations
taken in the supernova rest frame is given by

\begin{eqnarray}
  K_{xy} & = &  -2.5 \log
    \left(
    \frac
       {\int {\cal Z}(\lambda)S_x(\lambda)d\lambda}
       {\int {\cal Z}(\lambda)S_y(\lambda)d\lambda}
    \right)
    +2.5 \log(1+z) \nonumber \\
    & & \mbox{} ~~~+2.5 \log
    \left(
    \frac
        {\int F(\lambda)S_x(\lambda)d\lambda}
        {\int F(\lambda/(1+z))S_y(\lambda))d\lambda}
    \right)\nonumber \\
    & = & -2.5 \log
    \left(
    \frac
       {\int {\cal Z}(\lambda)S_x(\lambda)d\lambda}
       {\int {\cal Z}(\lambda)S_y(\lambda)d\lambda}
    \right) \nonumber \\
    & & \mbox{} ~~~+2.5 \log
    \left(
    \frac
        {\int F(\lambda)S_x(\lambda)d\lambda}
        {\int F(\lambda')S_y(\lambda'(1+z))d\lambda'}
    \right)
\label{Kij}
\end{eqnarray}
where
$F(\lambda)$ is the spectral energy distribution at the supernova, 
$S_x(\lambda)$ is the $x$'th filter transmission, and
${\cal Z}(\lambda)$ is an idealized stellar spectral energy distribution
at $z=0$ for which $U=B=V=R=I=0$ in the photometric system being used.
$K_{xy}$ is thus defined so that $m_y=M_x+\mu+K_{xy}$.  If
$S_x = S_y$, the first term drops out and this reduces to
the $K$ correction of Oke \& Sandage.  If $S_x(\lambda)$ is proportional
to $S_y(\lambda(1+z))$, then statistical and systematic errors
from the SN spectra, through the second term, are reduced.

We calculate generalized K corrections using Equation~\ref{Kij}
with Bessell's (1990) \nocite{be:ubvri} color zeropoints and
realizations of the Johnson-Cousins UBVRI
filter system.  (Note that Hamuy et al. (1995) had previously calculated
single-filter standard K corrections.)
We use a sample of SN data that includes 29 spectra from epochs
$-14 \le t_{max}^{B} \le 76$
days (in the supernova rest frame)
after blue maximum for SN 1981B, SN 1990N, and SN 1992A.
The SN 1981B data are from Branch et al. (1983)\nocite{br:sn1981b}, SN 1990N
data are described in
Leibundgut et al. (1991), and SN 1992A data are described in
Suntzeff et al. (1995)\nocite{su:sn1992}
and Kirshner et al. (1993)\nocite{ki:sn1992a}.

\begin{figure}
\psfig{figure=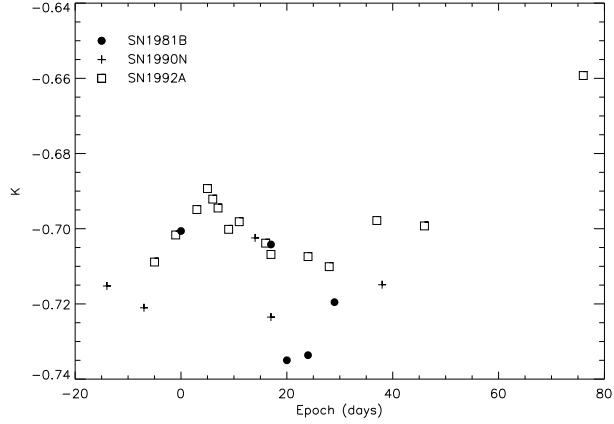,height=6cm}
\caption[figurecaption]{
  $K_{BR}(z=0.5)$ as a function of epoch for SN 1981B, SN 1990N, and SN 1992A.}
  \label{kbr5}
\end{figure}

\begin{figure}
\psfig{figure=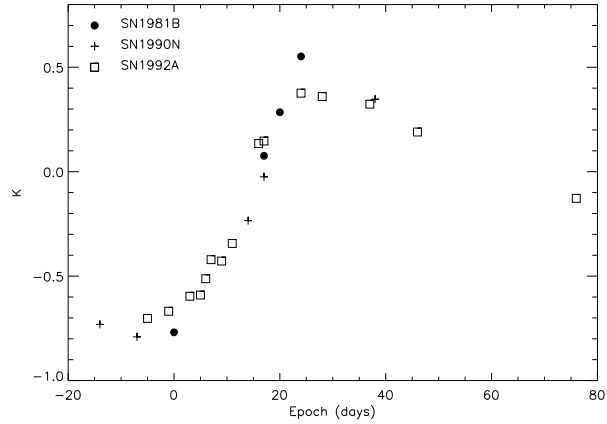,height=6cm}
\caption[figurecaption]{
  $K_{VR}(z=0.5)$ as a function of epoch for SN 1981B, SN 1990N, and SN 1992A.}
  \label{kvr5}
\end{figure}

Figures~\ref{kbr5} and \ref{kvr5} plot $K_{BR}$ and $K_{VR}$  for $z=0.5$
as a function of SN rest frame epoch.  The scatter in the data points
reflects both statistical
errors in the spectra, as well as SN to SN differences in K corrections.  They
also demonstrate the reduction of errors when filter pairs are chosen to match
at the appropriate redshift, i.e. $S_x(\lambda)$ proportional to
$S_y(\lambda(1+z))$.  Based on analysis of different filter combinations
at a range of redshifts, we find that with proper filter choice, these errors
can be constrained within $<0.05$ mag for redshifts $0<z<0.7$ and to
within even smaller
errors at epochs prior to 20 days after maximum light.

For more detailed discussion of the generalized K correction, see Kim,
Goobar, \& Perlmutter (1996)\nocite{kim:1996}.

\section{Hubble Constant Constant?}

We consider objects at a distance of $z > 0.3$ to be within the
cosmological flow of standard Friedmann cosmology.  Thus, the expected
peak apparent magnitude of a Type Ia supernova beyond $z=0.3$  can be calculated
as a function of the mass density of the universe $\Omega_M$
($\Lambda=0$):
\begin{equation}
m_R=M_B+5\log(3\times10^3 R_L(z;\Omega_M))+K_{BR}+25-5\log(h_0),
\end{equation}
where $K_{BR}$ is the K correction and
\begin{equation}
R_L(z;\Omega_M) =
\frac{4}{\Omega_M^2}\left[1+\Omega_M(z-1)/2+(\Omega_M/2-1)\sqrt{\Omega_Mz+1}\right]
\label{R}
\end{equation}
If we use $M_B = -18.52 +5\log(h_L/0.85) \pm 0.06$ from
Vaughan, Branch, \& Perlmutter 1995\nocite{va:95}, we find
\begin{equation}
5\log\left(\frac{H_L}{H_0}\right)=m_R-5\log(R_L)-K_{BR}-24.22 \pm 0.06.
\label{answer}
\end{equation}

\begin{figure}
\psfig{figure=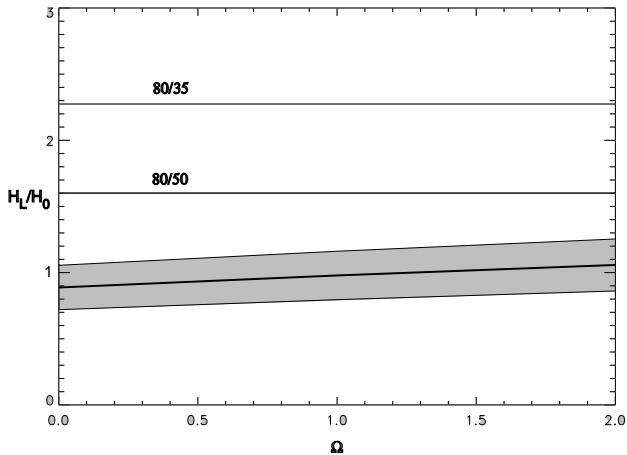,height=6cm}
\caption[figurecaption]{
  $H_L/H_0$ as determined from seven supernovae, as a function of
  $\Omega$ ($\Lambda=0$).}
  \label{hconst}
\end{figure}

We have used preliminary analysis of seven high redshift supernovae
(described by Perlmutter et al. in this volume), to calculate
values for $H_L/H_0$ for a range of $\Omega_M$, varying from 0 to 2
(see Figure~\ref{hconst}).
The shaded region represent a one sigma error bar on the value of
$H_L/H_0$.
A value of $H_L/H_0=80/35=2.29$ is strongly excluded, making peculiar
velocities an unlikely cause for the disparity between Freedman et al.'s
(1994)\nocite{fr:hubble} recent $H_0$ 
measurements
for Virgo Cepheids and theoretical arguments (e.g. Bartlett et al.) for
a low Hubble constant.
We reach the same conclusion when using the magnitude corrections
based on the width-brightness relation described in Perlmutter et al. 1996.

With the current data, it is impossible to simultaneously measure $q_0$
and $H_L/H_0$.  In the future, with SNe well sampled
in a broad range of redshifts, it should be possible to look for
significant spatial (or temporal) deviations from the predicted
Friedmann expansion.

\section{Conclusions}
We conclude that filter-matching K corrections give small enough errors
to make SNe useful as cosmological probes.
In the future, we will calculate K corrections using more SN spectra and
we will search for
a relation between light curve shapes and
K corrections.  Although we do not see 
such a correlation for the SNe that we have examined so far, which
have only moderate departures 
from the Leibundgut template lightcurve,
such a relation is likely to exist at some level since
SN colors appear to be correlated with light curve shape, as discussed
at this meeting by Riess et al.(1996) and Suntzeff et al.(1996).

We find that 
we can rule out significant spatial variation of the Hubble
constant for $0 \le \Omega_M \le 2$.
We plan to 
calculate $H_L/H_0$ for cosmologies with a cosmological constant,
although
positive values of $\Lambda$ will only strengthen our limit.

This work was supported in part by the National Science Foundation
(ADT-88909616) and
U.~S. Department of Energy (DE-AC03-76SF000098).

\bibliography{../alex}

\begin{thebibliography}{}

\bibitem{ba:h30}
J.~G. Bartlett, A.~Blanchard, J.~Silk, and M.~S. Turner.
\newblock {\em Science}, 267:980--983, 1995.

\bibitem{be:ubvri}
M.~S. Bessell.
\newblock {\em PASP}, 102:1181, 1990.

\bibitem{br:sn1981b}
D.~Branch, S.W. Falk, M.L. McCall, P.~Rybski, A.K. Uomoto, and B.J. Wills.
\newblock {\em ApJ}, 270:123, 1983.

\bibitem{br:araa}
D.~Branch and G.~A. Tammann.
\newblock {\em ARAA}, 30:359, 1992.

\bibitem{fr:hubble}
W.~L. Freedman, B.~F. Madore, J.~R. Mould, R.~Hill, L.~Ferrarese, R.~C.
  Kennicutt, A.~Saha, P.~B. Stetson, J.~A. Graham, H.~Ford, J.~G. Hoessel,
  J.~Huchra, S.~M. Hughes, and G.~D. Illingworth.
\newblock {\em Nature}, 371:757, 1994.

\bibitem{ha:hubble}
M.~Hamuy, M.~M. Phillips, J.~Maza, N.~B. Suntzeff, R.~A. Schommer, and
  R.~Aviles.
\newblock {\em AJ}, 109:1, 1995.

\bibitem{kim:1996}
A.G. Kim, A.~Goobar, S.~Perlmutter. 
\newblock {\em PASP}, In Press, 1996.

\bibitem{ki:sn1992a}
R.P. Kirshner and {et al.}
\newblock {\em ApJ}, 415:589, 1993.

\bibitem{nu:1995}
P.~Nugent, M.~Phillips, E.~Baron, D.~Branch, and P.~Hauschildt.
\newblock {\em ApJ}, 455L:147, 1995.

\bibitem{ok:kcorr}
J.~B. Oke and A.~Sandage.
\newblock {\em ApJ}, 154:21, 1968.

\bibitem{sn1992bi}
S.~Perlmutter, C.~R. Pennypacker, G.~Goldhaber, A.~Goobar, R.~A. Muller,
H.~J.~M. Newberg, J.~Desai, A.~G. Kim, M.~Y. Kim, I.~A. Small, B.~J. Boyle,
C.~S. Crawford, R.~G. McMahon, P.~S. Bunclark, D.~Carter, M.~J. Irwin, R.~J.
Terlevich, R.~S. Ellis, K.~Glazebrook, W.~J. Couch, J.~R. Mould, T.~A. Small,
and R.~G. Abraham.
\newblock {\em ApJ}, 440:L41, 1995.

\bibitem{pe:spain}
S.~Perlmutter et al.
\newblock {This Volume} 1996

\bibitem{re:lcs}
A.~G. Riess, W.~H. Press, and R.~P. Kirshner.
\newblock {\em ApJ}, 438:L17, 1995.

\bibitem{re:spain}
A.~G. Riess et al.
\newblock {This Volume} 1996

\bibitem{su:sn1992}
N.B. Suntzeff.
\newblock {In preparation}, 1995.

\bibitem{su:spain}
N.B. Suntzeff et al.
\newblock {This Volume} 1996

\bibitem{tu:hubble}
E.L. Turner, R.~Cen, and J.~P. Ostriker.
\newblock {\em ApJ}, 103:1427, 1992.

\bibitem{va:inp}
T.~Vaughan, D.~Branch, D.~Miller, and S.~Perlmutter.
\newblock {\em ApJ}, 439:558, 1995.

\bibitem{va:95}
T.~Vaughan, D.~Branch, and S.~Perlmutter.
\newblock {\em ApJ submitted}, 1995.

\end{thebibliography}
\bibliographystyle{plain}
\end{document}